\documentclass[journal,10pt,twocolumn,twoside]{IEEEtran}
\usepackage{mathtools}
\usepackage{empheq}
\usepackage{algpseudocode, algorithm}
\usepackage{algorithmicx}
\usepackage{subfig}
\usepackage{varwidth}
\usepackage{url}
\usepackage{amssymb}
\usepackage{mathtools}
\usepackage{changes}
\usepackage[short]{optidef}
\usepackage{caption}
\usepackage{amsmath}
\usepackage{textcomp}
\usepackage{gensymb}
\usepackage{cite}
\usepackage{booktabs}
\usepackage{hyperref}
\usepackage{balance}

\begin{document}

\title{Towards Blockchain for {Edge-of-Things}: A New Paradigm, Opportunities, and Future Directions
}

\author{Prabadevi B, N Deepa, Quoc-Viet Pham, Dinh C. Nguyen, Praveen Kumar Reddy M, \\ Thippa Reddy G, Pubudu N. Pathirana,~\IEEEmembership{Senior Member,~IEEE,} and Octavia Dobre,~\IEEEmembership{Fellow,~IEEE} }

\IEEEtitleabstractindextext{%
\begin{abstract}
Blockchain is gaining momentum as a promising technology for many application domains, one of them being the {Edge-of-Things (EoT)} that is enabled by the integration of edge computing and the {Internet-of-Things (IoT)}. Particularly, the amalgamation of blockchain and EoT leads to a new paradigm, called {blockchain enabled EoT (BEoT)} that is crucial for enabling future low-latency and high-security services and applications. This article envisions a novel BEoT architecture for supporting industrial applications under the management of blockchain at the network edge in a wide range of IoT use cases such as smart home, smart healthcare, smart grid, and smart transportation.  The potentials of BEoT in providing security services are also explored, including access authentication, data privacy preservation, attack detection, and trust management. Finally, we point out some key research challenges and future directions in this emerging area. 
\end{abstract}

}

\markboth{Accepted at the IEEE Internet of Things Magazine}%
{}

\maketitle

\IEEEdisplaynontitleabstractindextext

\IEEEpeerreviewmaketitle

\section{Introduction}

\IEEEPARstart{T}{he} {Internet-of-Things (IoT)} is the interconnection of "things" which are connected to the Internet. Due to the advancements in information and communication technology and the affordability of the Internet, \textcolor{black}{a} number of IoT-based applications has been on the rise in recent years. The rapid developments of smart cities and human-centric services across the globe have contributed to the increase in the {IoT-based} applications. {To handle the voluminous data generated from the IoT devices at regular intervals, most IoT applications have relied on cloud computing for data processing and storage. However, this cloud-based model suffers from high communication overhead and thus is not suitable for time-sensitive IoT applications. 
}

{On the other side, multi-access edge computing (MEC) has emerged as a promising technology to support IoT systems by allocating computation and storage resources at the network edge for low-latency and real-time services \cite{porambage2018survey}. As a result, the integration of MEC with IoT creates a new model as {Edge-of-Things (EoT)} \cite{el2017edge}. In EoT, MEC nodes are placed in near proximity of the IoT devices for storage and low-latency computation, making it well \textcolor{black}{suited} for performing real-time data tasks in IoT applications
such as self-driving cars, automatic manufacturing, and object detection \cite{porambage2018survey}.}
\begin{figure*}[t]
	\centering
	\includegraphics[width=0.925\linewidth]{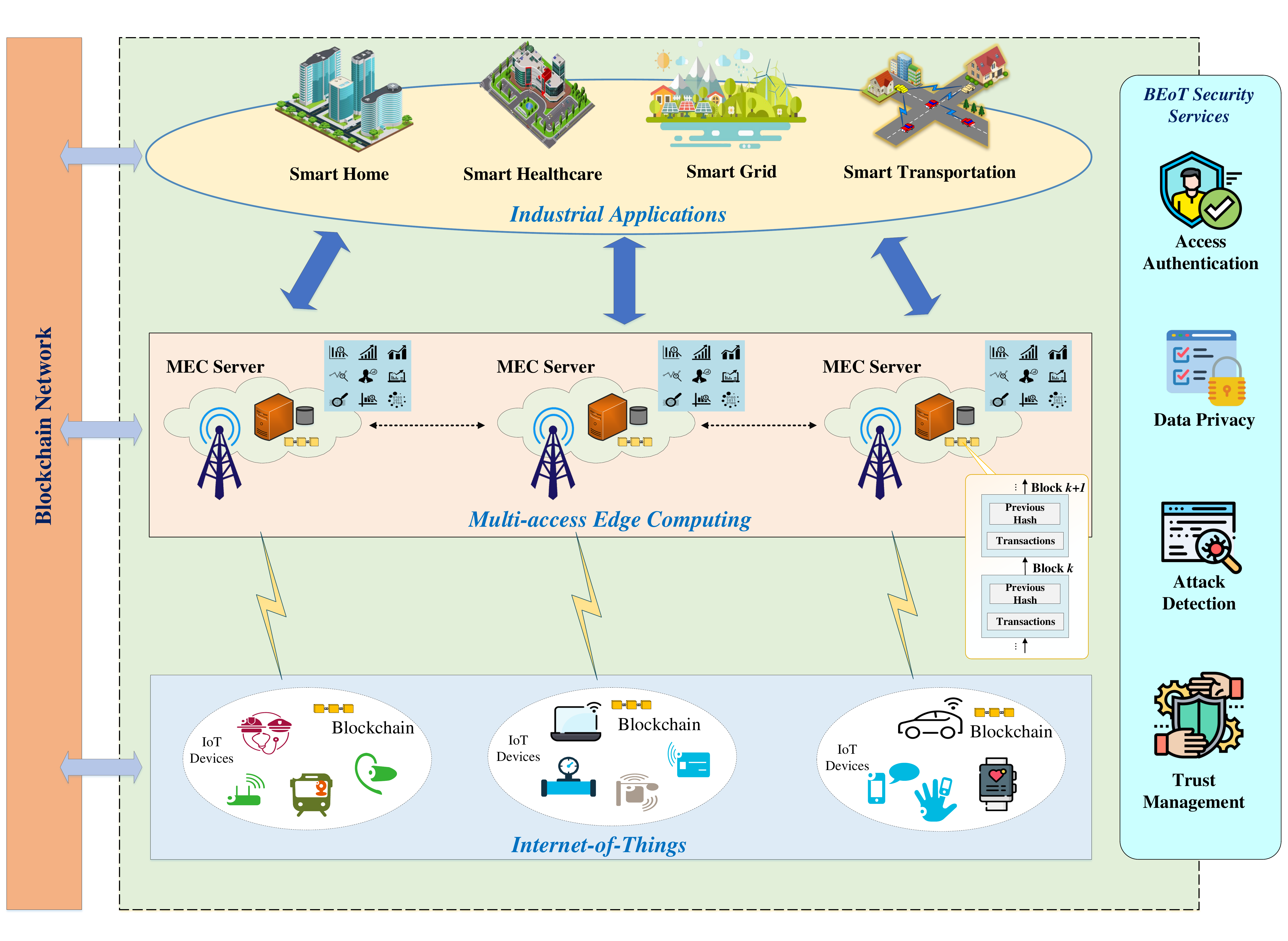}
	\caption{The generic architecture of BEoT.}
	\label{Fig1}
\end{figure*}

\textcolor{black}{Although EoT has been widely adopted in various customer applications such as smart transportation and smart healthcare for improving user quality-of-experience, some critical challenges concerned with security and data distribution at edge needed to be explored. 
For example, healthcare systems with mobile devices generate sensitive user data such as personal data and personal preferences which may possess an elevated risk of compromised security. The blockchain technology with its unique features such as traceability, immutability, and decentralization makes it an ideal solution to provide high security for EoT applications and networks. This is enabled by its network architecture, where each block in the blockchain network stores data transactions and the hash of the previous node, creating a chain of blocks that is maintained by a consensus mechanism such as proof-of-work (PoW). These properties of the blockchain ensure that the records are secure and tamper-proof. Particularly, the convergence of blockchain and EoT form a new paradigm called blockchain-enabled EoT (BEoT) which supports low-latency data services with degrees of security and privacy for IoT applications such as smart grids, smart healthcare, and smart homes \cite{7}.}

\textcolor{black}{The use of blockchain would provide the following key new features for EoT networks:
\begin{itemize}
    \item The immutability and traceability features of blockchain can be leveraged to ensure the reliability of the transactions in industrial applications such as smart grid, smart transportation, smart health care, etc. 
   \item The consensus mechanism of the blockchain guarantees the trustworthiness and transparency of information transferred over the BEoT network.
   \item Moreover, the decentralization of blockchain has the potential to ensure low-latency response for EoT networks when compared to the traditionally centralized network architectures, which would facilitate customer services and applications. 
\end{itemize}}

{In this article, we propose a novel BEoT architecture, 
as illustrated in Fig.~\ref{Fig1}. The proposed architecture consists of three main entities, \textcolor{black}{namely} {IoT}, MEC, \textcolor{black}{and} blockchain, along with industrial applications and BEoT security services. }
{
\begin{itemize}
	\item \textit{IoT:} IoT devices are responsible for generating or gathering data from the physical environments and then \textcolor{black}{transmitting them} to the nearby edge servers via access points or base stations. IoT devices with certain resources (e.g., smart phones and laptops) can act as a mobile blockchain entity to make transactions to communicate directly with the MEC servers, while lightweight IoT devices can participate in the blockchain network via their representative gateways (e.g., mobile phones).  
	\item \textit{MEC:} In BEoT networks, MEC servers can offer computing and storage resources to handle data tasks offloaded from IoT devices and provide data services for end users, ranging from data analytics, {to data mining, \textcolor{black}{to} prediction and storage}. Moreover, each MEC server can also act as a blockchain miner to perform block mining for maintaining the blockchain network. 
	\item \textit{Blockchain:} Blockchain is to form the BEoT system running on top of the EoT network, aiming to interconnect IoT devices, MEC servers and end users together in a decentralized fashion. Particularly, blockchain ensures reliable operations of BEoT systems without requiring any third party by using its inherited services such as data consensus, smart contracts, and shared ledgers \cite{3}. \textcolor{black}{More specifically, data consensus provides verification services on user transactions by using mechanisms such as PoW managed by a network of miners. This service is highly necessary for BEoT in improving blockchain consistency and ensuring high network security. Smart contracts can provide self-executing and independent features to build business logic and trust in the BEoT system. They also provide security services on user access authentication or data sharing verification once the IoT peer nodes perform transactions, which also supports maintaining security over the edge blockchain. Moreover, shared ledger represents the database that is shared and distributed among BEoT members (e.g., IoT devices, MEC servers and end users). The shared data ledger records transactions, such as information exchange or data sharing among IoT devices and MEC servers. It enables industrial networks where users can control and verify their own transactions when communicating with the MEC server.}
	\item \textit{Industrial applications:} BEoT can enable new industrial applications. 
	For example, in a BEoT-based smart transportation system, MEC servers can support low-latency traffic analytics, while blockchain ensures secure vehicular communications among distributed vehicles \cite{dai2020deep}. 
	\item \textit{Security services:} Enabled by the inherent security properties such as decentralization, immutability, and traceability, blockchain can provide a number of important security services for BEoT, including access authentication, data privacy, attack detection, and trust management. The analysis of such security services will \textcolor{black}{subsequently be presented in detail.} 
\end{itemize}}

Motivated by the recent research efforts in blockchain and EoT \cite{7}, this article presents a more genertic BEoT architecture which capture the most important features and applications of BEoT in industrial domains.  Enabled by the proposed BEoT architecture, in this article, we provide an overview of the use of BEoT in many industrial applications. Then, we explore the security opportunities brought by blockchain for BEoT networks and services. We conclude with a discussion of some important technical challenges and future propositions for BEoT research. 

\section{{BEoT for Security Services}}
This section presents BEoT's interventions to address security and privacy in smart cities. Due to exponential growth in technology, cybersecurity and efficient data sharing are primary concerns for organizations and individuals.BEoT can provide effective and reliable information exchange among the individuals as the blockchain offers secure transparent transactions that can be monitored continuously.
\subsection{Authentication and Authorization} 
\begin{figure}[t]
	\centering
	\includegraphics[width=0.925\linewidth]{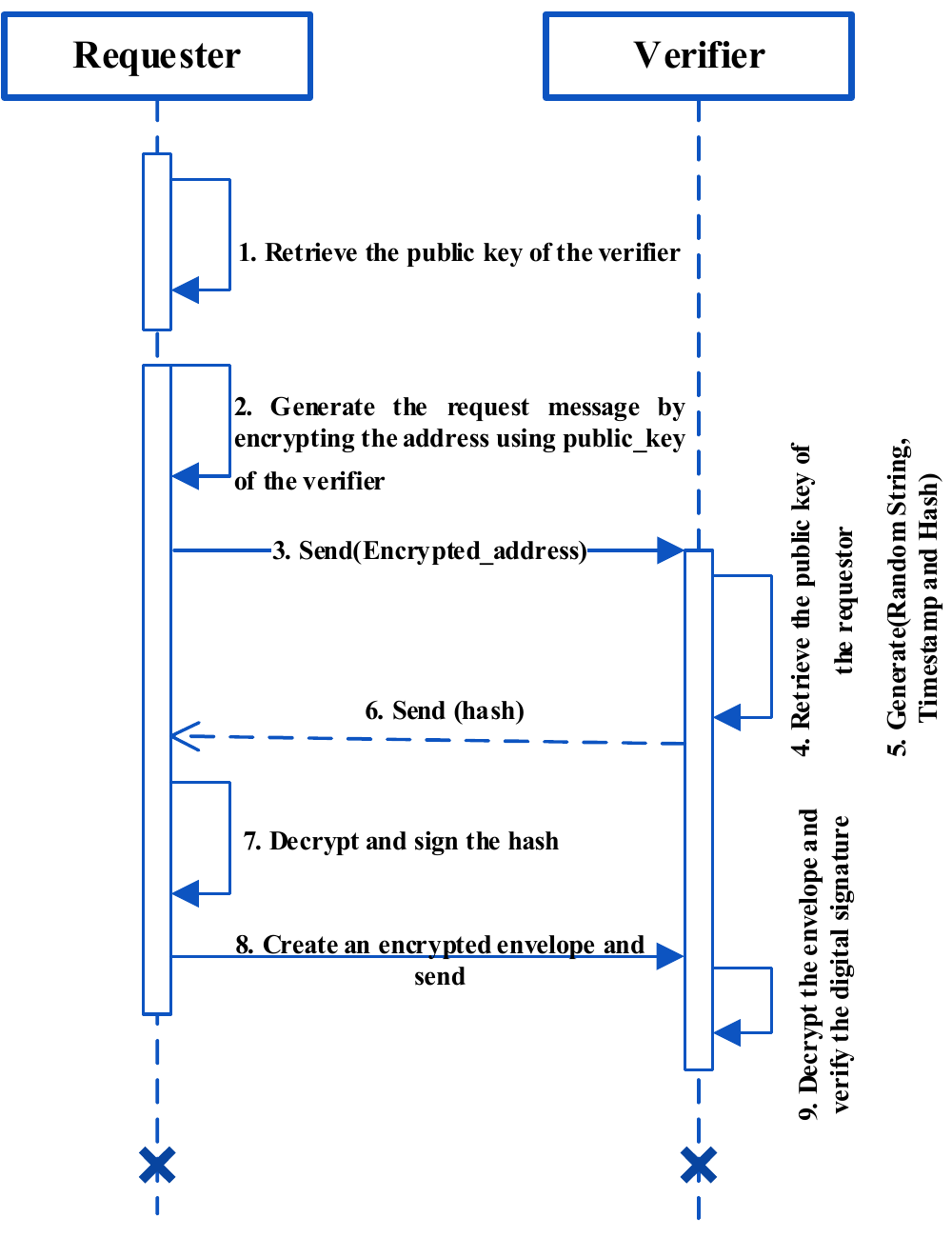}
	\caption{\textcolor{black}{Blockchain-Based Authentication Sequence Diagram.}}
	\label{Fig:Sequence}
\end{figure}

\begin{figure*}[t]
	\centering
	\includegraphics[width=.85\linewidth]{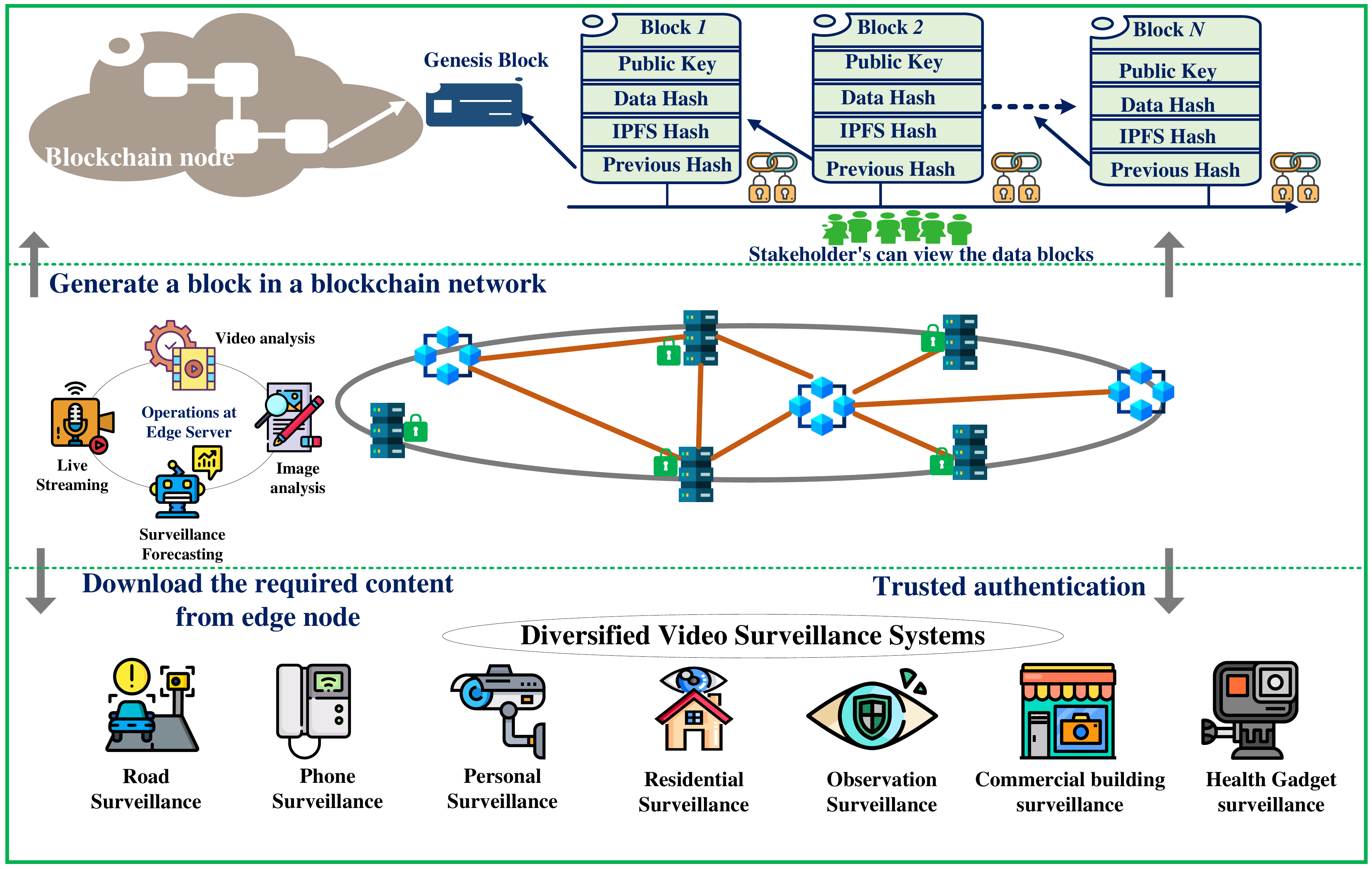}
	\caption{ BEoT-based secure authentication in surveillance data sharing for industrial applications.}
	\label{Fig:BEoT_safesurveillance_datasharing}
\end{figure*}

The data generated by IoT devices has become progressively complex and nuanced, having a significant impact on the reliable and efficient transmission of data to the end user. It is, therefore, vital to ensure efficient authentication and authorization between the various key enabling technologies. \textcolor{black}{Fig.~\ref{Fig:Sequence} depicts the sequence diagram for Blockchain-based authentication model.} However, the convergence of BEoT can meet the standards for the better control of these types of authentication and authorization challenges. \textcolor{black}{Byzantine tolerance consensus approaches are used to build a blockchain to ensure secure communication, authentication and end-to-end quality control. Resolution edge nodes and cache nodes are deployed at the edge, offering edge authentication, and boosting the hit ratio \cite{11}.} BEoT pays a lot of attention to facilitate mutual authentication for smart grid applications. The key agreement protocol is used at the edges where the edge server information is added to the blockchain, thus avoiding the leakage of the smart meter information. Using a smart contract allows only registered users to be connected to the public key, which helps to maintain the security and reliability of the smart grid network. \textcolor{black}{Smart contract processing of key inputs can be used to reinforce the key  revocation thus, preventing the need to prefer a trusted centre and preventing a design flaw.} In VEC, authenticating and authorizing information exchange is a challenge, as vehicle networks are dynamic. BEoT promises secure, efficient transmission, trackable map paths that use a dynamic decentralized route hash chain and ensure a reliable system with low communication overhead. \textcolor{black}{Blockchain technology integrated with the 5G RFID supply chain system has facilitated reliable computing and storage costs for EoT. The authentication mechanism uses the cryptographic hash function and bitwise XOR rotation, which consists of N blocks, and each block has a reader tag. The reader tag confirm its authenticity followed by an acknowledgement of the requested data.}
{Moreover, BEoT provides efficient authentication for surveillance data sharing, an important service in industrial applications, as indicated in Fig.~\ref{Fig:BEoT_safesurveillance_datasharing}.}

\subsection{Data Privacy}
In any surveillance application, resource utilization, maintaining low latency, and privacy-protection are challenging tasks. \textcolor{black}{Blockchain, in combination with the Nudge Theory, breaks down the data disclosure schemes of new customers with the collaborative customer. Filtering framework plays a key role in ensuring the privacy of task scheduling and access management for edge devices.} 
{BEoT's} privacy mechanism is accomplished in {industrial IoT} applications through smart contracts where each \textcolor{black}{node's} task information can be monitored. All the available edge nodes are interconnected to a distributed platform and the data is scattered using the alias feature of blockchain \cite{gai2019differential}. Edge devices need to carry out the required task and determine time and energy utilization. Power depletion and electrical grid security flaws are key risks in smart grid applications. BEoT overcomes inappropriate users by monitoring \textcolor{black}{the illegal consumption of} electricity. Network nodes with high computational capability (e.g., MEC servers), are configured in the blockchain, taking responsibility for secure resource allocation,  using {an} authorization scheme. Edge nodes supply power to a valid end-user that saves costs on a centralized server and aims to improve the computing process. The information obtained and produced {by vehicles} mostly consists of sensitive data. Limited memory size and data privacy are \textcolor{black}{challenges} for vehicle networks. BEoT seeks to overcome these challenges by providing a blockchain consortium that focuses on delivering data privacy and effective information sharing across {vehicular networks\cite{9}}. Blockchain provides a reputation-based mechanism to edge nodes that maintain vehicle reputation details when interacting between one vehicle and another, thus enhancing data reliability.
\subsection{Security Vulnerabilities}
With the incredible expansion of communications technology, malicious hackers explores new security vulnerabilities to leverage highly sophisticated threats, which are extremely difficult for frameworks to prevent. {Software defined networking (SDN)}-enabled blockchain technology detects malicious attacks in the cloud, improves storage space and significantly improves latency at the edge.  The blockchain platform is used to avoid malicious attackers in {economic denial of sustainability}. \textcolor{black}{The secret sharing system is applied to provide protection when the edge server fails. However, sometimes, when an attacker targets the edge server, the binary search method is used to analyze and trace the edge server that is affected.} Vehicle edge computing using blockchain \textcolor{black}{can be used} to deliver event-driven messages (EDMs) to end users to optimize communication costs and security attacks. EDMs are delivered to proximity edge servers to minimize processing time. Blockchain uses smart contract at the edge to monitor responses and prevent attacks from occurring. The {blockchain} system guarantees a decentralized distributed caching system in which the automobiles enable efficient caching and facilitate an authorized block at the nearest base stations, \textcolor{black}{preventing} the attackers from carrying out malicious activities. Blockchain uses proof-of-utility (PoU) computations to strengthen the block validation process \cite{dai2020deep}. {Hence, blockchain can be effectively used to address security vulnerabilities in EoT.}

\subsection{Trust Management}
MEC offers a variety of services to increase the efficiency of data transfer and communication overhead and, therefore, to ensure end-to-end trust, while the blockchain technology has been used to build trust at the edge of the network. \textcolor{black}{A blockchain-based, trusted data management system (BlockTDM) has been proposed in \cite{zhaofeng2019blockchain} to provide edge node data protection in IoT networks. BlockTDM offers data security and privacy through the use of a multi-channel data segment. Before the data is processed inside the blockchain, encryption techniques are used to ensure trust management.} Hyperledger pays special attention to perform decryption and secure data transactions in a blockchain. Blockchain combines with a Reinforcement learning (RL) algorithm to overcome computational latency, streamline power usage, and load balancing of edge nodes.
\textcolor{black}{The RL algorithm specifies the number of edge processors to complete the computational tasks, depending on the size of the task and the blockchain messages.}
An Ethereum system is also integrated to provide a federated, secure edge computing system in which blockchain tokens are generated for reliable transactions. {From the above discussion, it is clear that blockchain provides trust management for EoT.}

\section{{BEoT for Industrial Applications}}
\begin{figure*}[t]
	\centering
	\includegraphics[width=0.90\linewidth]{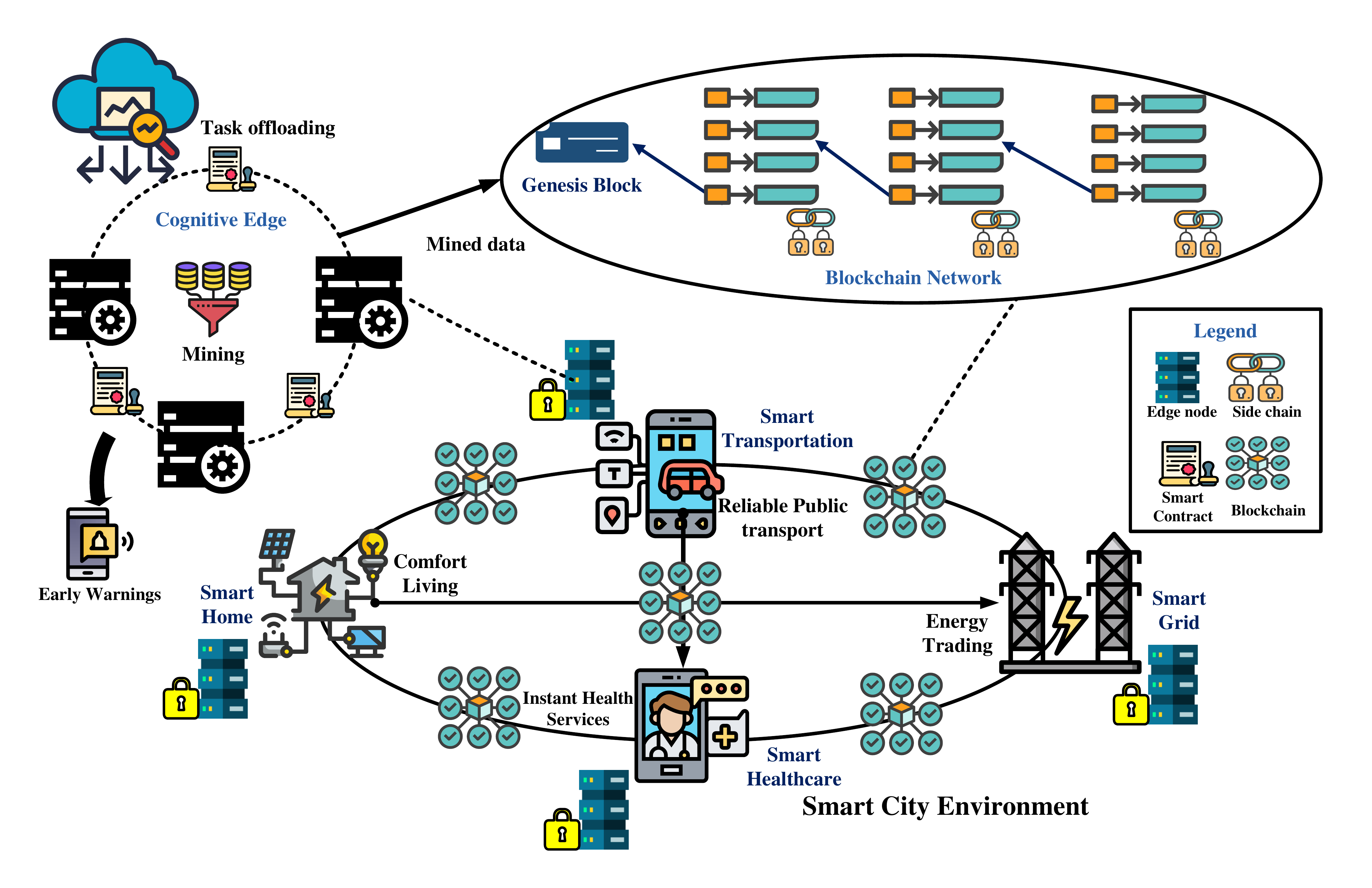}
	\caption{{Secured transactions in BEoT-based industrial applications.}}
	\label{Fig2}
\end{figure*}

{This section presents the use of BEoT in some key industrial application domains, namely smart home, smart healthcare, smart grid, and smart transportation, as shown in Fig. \ref{Fig2}. }


\subsection{BEoT in Smart Home}
{Nowadays, it is mandatory to provide a secure and safe environment for elderly and disabled people. 
Blockchain can \textcolor{black}{provide} a viable solution to protect all IoT devices in smart home and data acquired from these devices for establishing secure communication using immutable ledgers among entities such as data servers, smart home devices, and {homeowner \cite{7}}. Further, blockchain can provide reliable authentication to the devices and prevent data theft by using smart contracts that can perform user access verification at the smart home gateway. Additionally, IoT devices are exceptionally delicate, and the maximum amount of data is transmitted to the cloud environment in a remote location for processing. It is challenging to provide scalability for the centralized cloud environment where a \textcolor{black}{large} amount of data are generated by ubiquitous devices, and IoT devices and cloud are located far away from each other. \textcolor{black}{In this context, MEC would be a highly effective technique to enhance the networking, computing, intelligent and storage facilities for IoT devices in smart home settings, with the ability to provide faster response and reduce network traffic at the network edge as specified in the BEoT architecture.}}.

\subsection{BEoT in Smart Healthcare}
{The advancement of artificial intelligence (AI), IoT, mobile computing, cloud computing and big data, the conventional healthcare system is converted to smart health services. 
People can monitor their health conditions using wearable devices and obtain medical assistance from online systems. Several issues such as privacy, security, interoperability, transparency, data storage and decentralization are being faced during the implementation of \textcolor{black}{the} centralized smart healthcare service. Blockchain with MEC helps transform the current centralized health architecture to a more decentralized and secure system. It enables to synchronize the medical data collected, provides data security using encryption, authenticates the users and allows data storage in the server. {Blockchain integrated with MEC includes security, data sharing and decentralization for medical data where the patient is allowed to share the data with trusted members.} The MEC processes the medical data and overcomes high bandwidth problems. In some other smart healthcare systems, blockchain with MEC provides access control and protects the access events. A mutual agreement facility is applied to all the access events for verification and storage. The medical data is saved in edge nodes, and access control policies are applied by the {edge nodes} to obtain attribute-based access using blockchain. The authorized users with attribute-based access control are allowed to access the medical data. \textcolor{black}{Thus, in the proposed BEoT architecture, the healthcare data is collected through IoT devices and shared using blockchain within the MEC framework in order to provide security, transparency and privacy for the smart healthcare system.}}

\subsection{{BEoT in Smart Grid}}
Due to the population growth, the electricity demand also increases. The smart grid offers effective trading of electrical energy in a smart city environment. 
The decentralization and distributed nature of smart grids facilitate the efficient utilization of power resources. 
Energy trading through the smart grid is prone to numerous security attacks through which the attackers try to deceive the consumers' usage \textcolor{black}{profiles} and their electricity. Thus, the blockchain framework with assisted cryptographic techniques can be utilized for secured energy transactions without data leakage, and smart contracts enabled MEC services can be used for attaining {low-latency response \cite{10}.} {Moreover, the usage of various cryptographic techniques with BEoT should not incur a high computational overhead; therefore, a lightweight consortium, blockchain was recommended for better scalability and effective energy trading. \textcolor{black}{In the BEoT architecture, blockchain integrated with data aggregation facility can be applied in the smart grid for privacy-preserving transactions. MEC nodes are responsible for decisions on energy trading in order to minimize the processing time, which leads to the reduction of latency; whereas also blockchain provides security for transactions in energy trading. Blockchain has been successfully deployed in many applications such as energy Internet, solar electricity exchanges and energy auctions.}}

\subsection{{BEoT in Smart Transportation}}
recent advancements in satellite communication, vehicular technologies and IoT, intelligent transport system (ITS) has become an emerging research area during recent years. 
The conventional centralized system faces many issues like security, data storage, {and server failure}.  The implementation of blockchain in the \textcolor{black}{Internet of Vehicles(IoV)} platform has been explored to support the information exchange necessity of ITS \cite{liu2019electric}. The application of blockchain in {ITS provides} immutability, transparency, security, automation and decentralization facilities. Decentralized IoV network of blockchain includes distributed entities such as vehicles, {humans,} and roadside nodes. These entities are allowed to operate autonomously. Blockchain helps IoV to remove the entities such as the central control, trusted intermediate parties, {administrators,} and central service manager. The blockchain network allows the participants to manage the transactions and vehicular services independently which \textcolor{black}{results in reduced operational cost.} Blockchain \textcolor{black}{uses} advanced cryptographic algorithms to provide security and privacy for IoV networks \cite{9}. \textcolor{black}{In the BEoT architecture, data alteration and tampering in IoV networks can be prevented by the immutability property of blockchain. AI-based connected autonomous vehicle framework trains the learning model in MEC systems and provides the obtained knowledge to the blockchain. This enables collective intelligence for connected vehicles and avoids large data transmission. Also, blockchain aids in securing the distributed trained models. Thus, connected vehicles have several advantages such as low latency response thereby increasing the network performance.}

\section{{Research Problems and Future Directions}}
\begin{figure*}[t]
	\centering
	\includegraphics[width=0.90\linewidth]{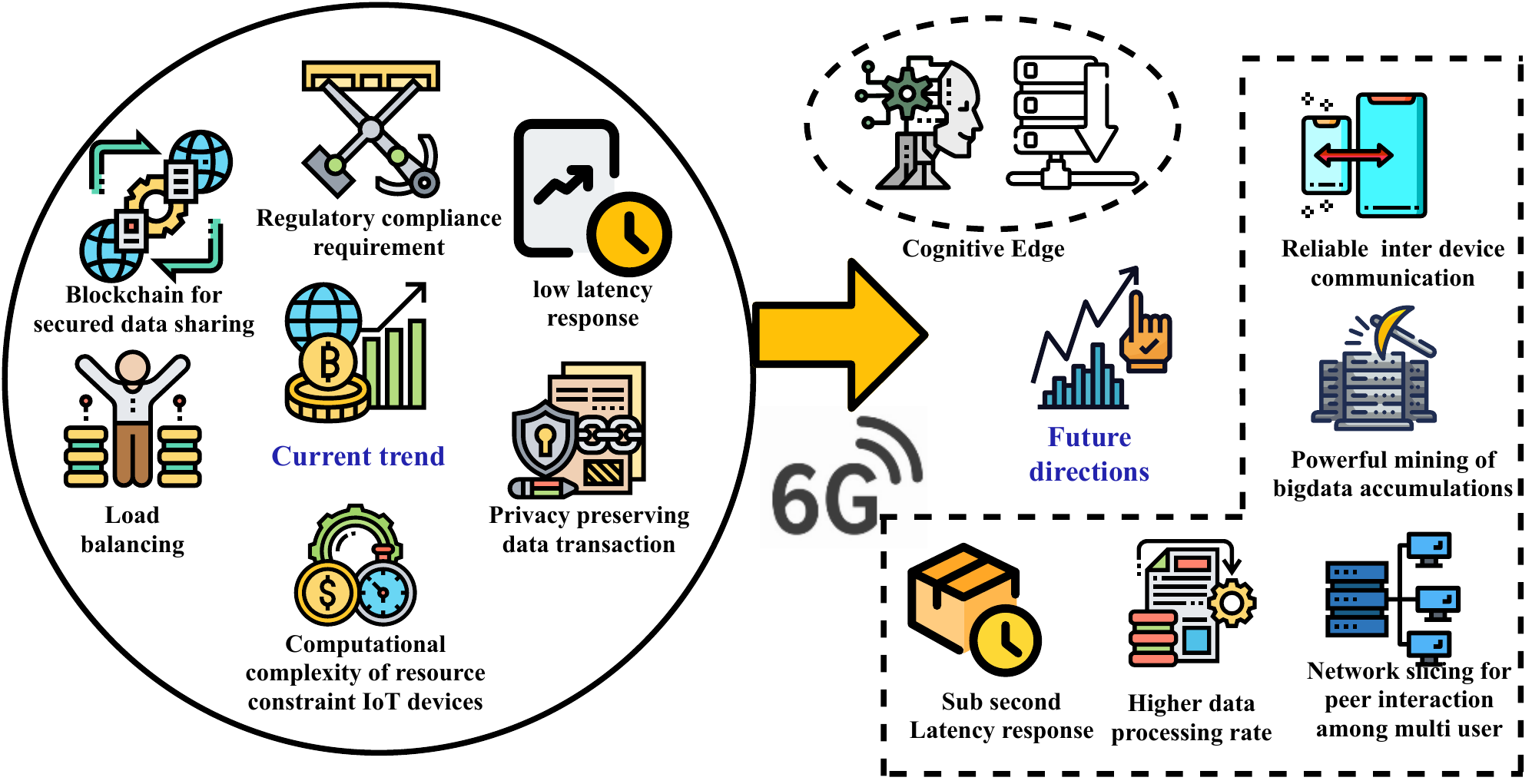}
	\caption{The curent trend and future directions of BEoT.}
	\label{Fig3:Direction}
\end{figure*}

The integration of blockchain with EoT \textcolor{black}{poses a plethora of research questions such as} secured data sharing, authorized access to diversified IoT devices, regulatory compliance requirements, standardized models for interoperability, load balancing, and resource management. This section provides the potential research problem of optimal resource utilization in the BEoT paradigm. Also, the future scope in the integration of blockchain with EoT is presented. \textcolor{black}{The current trends and future directions of BEoT \textcolor{black}{are} depicted in Fig.~\ref{Fig3:Direction}.}  

\subsection{Research Problems}
\subsubsection{Resource Utilization}
Resource management is a significant concern while integrating blockchain and edge computing with IoT. IoT devices are resource-restricted (i.e. low bandwidth) and have low storage capacity. Also, they incur more computational overhead while participating in the blockchain network as encrypting the blocks is mandatory. In contrast, the decentralized blockchain consumes more bandwidth and more storage. Real-time video streaming which consumes larger bandwidth is a mandated need for most of the applications in the smart city environment. Usually, the processing of data at blocks will consume more computational resources when the ledger is accumulated with more information. {As such}, side chains can be used with the blocks segregating the secondary data from the central ledger and storing them in the side chains \cite{4}. \textcolor{black}{Though the mining process in PoW consensus blockchains consumes more energy \cite{sedlmeir2020energy} as it is energy-intensive by design and has no significant impact when the blockchain scales up. Therefore, a lightweight blockchain with less energy consuming cryptocurrencies should be designed. Intel's recent work proves proof-of-the-elapsed-time (PoET) to leverage the trusted computing by enforcing random waiting times in block creation.} The computational resource utilization in data processing between the cloud and the end devices can be accompanied by edge servers to guarantee the low-latency response. This, in turn, makes the processing faster and ensures the availability of computational resources. The limited storage capacity at the edge cannot withstand the dynamic user demand, so the transactions at the edge can be managed using blockchains.  Furthermore, skyline queries that search only the related data instead of the whole database can be used for retrieving data information from massive data accumulation.  {Thus,} the effective resource utilization in the scalable BEoT \textcolor{black}{environments} alleviates the storage \textcolor{black}{overheads} involved in processing enormous data accumulation. {Also, it requires an authorized load provisioning system for sustainable services.}  

\subsubsection{Security in BEoT}

{
Although the BEoT paradigm can bring significant improvements in the performance of various smart systems, security still {remains a} considerable concern in this integration. \textcolor{black}{While} blockchain ensures privacy-preserving data transaction, it is prone to various security threats such as crypto key exchange, data leakage, regulatory needs, double spending of currency, linking transactions and handling data in the chain. The immutability of blockchains in the BEoT paradigm 
may open a loophole for third-party service providers to loot the data leading to various cyber-attacks. Also, existing security services such as information coding and digital signatures, impose a higher computational overhead. Further, those services shall introduce additional overhead in mining nodes which {perform both mining} and authorization. \textcolor{black}{The smart contracts or the chain code are written using different languages and are vulnerable to many security risks such as different interpretations for programming constructs and concurrency issues. Therefore, appropriate validation and verification must be carried for smart contracts to avoid vulnerable security risks. The open research challenges include privacy-preserving data transactions with lower computational overhead, the privacy of chain code, authorized third-party services, and segregating activities from the mining node to reduce the overhead. Above all, the new enforced techniques should be well-suited with the scalable nature of the BEoT paradigm.}}

\textcolor{black}{The proposed BEoT framework can alleviate the security issues concerned through its security services and ensures full trust in providing privacy-preserving data transactions among the IoT devices and MEC servers. The BEoT framework expedites all the services without relying on any third party, thereby leading to more secured and privacy-preserving data transactions. Though blockchain at MEC and IoT may consume more energy, BEoT can effectively distribute and manage its resources through faster data processing.}

\subsection{{Future Directions}}
\subsubsection{Cognitive Edge in BEoT}
{AI} uses the intelligent agents {that} can mimic the intelligence of the human brain without human {input. It is used to solve} complex decision problems by learning through previous experience. The application of these intelligent agents has bought remarkable advancements in {all domains.} Machine learning and deep learning, the subset of AI have significant contributions to these technological advancements.  AI's sophistication can be utilized for combating the computational challenges in the  BEoT environment. {Cognitive Edge (CE)} computing refers to the integration of AI's cognition with edge computing for effective processing at the edge \cite{5}. CE in BEoT environment allows the end devices to mine their data to be stored at blocks, thus alleviating the data processing at blocks and minimizing the bandwidth consumption for processing at blocks. 
\textcolor{black}{CE will assist in forecasting the load utilization through knowledge aggregation and ensure minimal data caching at the edge. Furthermore, for secured knowledge trading among edge nodes, blockchain-based credit coins can be used. As the smart city environment comprises a large number of heterogeneous devices, deep learning algorithms can be used for extracting meaningful information from these enormous data accumulations. Machine learning algorithms can be used for data representation and visualization.}
In the smart city environment, CE in BEoT will provide an effective resource utilization, fault prediction with minimal computational overhead, security attack classification, prediction and detection. Also, CE assists in load management with increased data processing rate. {Making the CE perform better in BEoT training the AI models in this diversified environment is still a challenging issue.} 

\subsubsection{Integration of BEoT with 5G and Beyond}
All the cellular networks are moving towards the adoption of faster \textcolor{black}{fifth-generation (5G) and beyond} systems which can carry a larger payload in a shorter duration with optimal resource utilization. 5G and beyond technologies can best host the application in {BEoT environment, as the BEoT environment} uses peer-to-peer communication for the transactions \textcolor{black}{with a} higher data processing rate. Due to the higher computational requirement for mining, blockchain utilizes the edge server for task offloading and low-latency response. BEoT in {5G provides faster} data processing with the low-latency response, global infrastructure sharing and network slicing to accommodate multiple users with peers interaction, as indicated in Fig.~\ref{Fig3:Direction}. The conjugal of 5G and CE with BEoT will embark the powerful mining of big data accumulations with a low-latency response. Beyond 5G, the sixth generation (6G) communications with the full support of AI and blockchain will be a better host for BEoT applications with CE. Consequently, while moving beyond 5G, the issues pertaining to 5G as well as underlying issues such as high data rate, upgraded {quality-of-services,} secured data transactions and low-latency response must be considered before adoption \cite{ chowdhury20196g}. 6G \textcolor{black}{will offer} sub-second latency with higher processing rates than 5G.
Therefore, the resource management constraint of BEoT can be effectively managed by the intelligence of cognitive edge and higher data processing of 5G \textcolor{black}{beyond} communication systems. 
 
\section{Conclusion}
In this article, we have introduced BEoT, a novel solution that \textcolor{black}{facilitates} low-latency computation of IoT applications while providing high security degrees at the network edge, empowered by the cooperation of MEC and blockchain. Specifically, we have explored the opportunities brought by BEoT in some key applied domains, namely smart home, smart healthcare, smart grid, and smart transportation. The roles of BEoT in providing security services have been also analyzed, ranging from access authentication, data privacy preservation to attack detection and trust management. Finally, we have pointed out some key research challenges and future directions. We believe that the significant improvements attainable over current BEoT approaches will pave \textcolor{black}{the} way for new innovative researches and solutions for enabling the next generation of BEoT. 

\section*{Acknowledgment}
The work of Quoc-Viet Pham (corresponding author) was supported by a National Research Foundation of Korea (NRF) Grant funded by the Korean Government (MSIT) under Grants NRF-2019R1C1C1006143 and NRF-2019R1I1A3A01060518.

\balance


\vskip -2\baselineskip 
\begin{IEEEbiographynophoto}{Prabadevi B} is an Assistant Professor (Senior) in School of Information Technology and Engineering, Vellore Institute Technology, Vellore, India, in 2018. She received her Ph.D. in Information Technology with Networking as a specialization from Vellore Institute Technology, Vellore. She is currently working in the areas of  Decision support systems and Machine Learning.  
\end{IEEEbiographynophoto}
\vskip -2\baselineskip 
\begin{IEEEbiographynophoto}{N Deepa}
is currently working as Assistant Professor (Senior) in School of Information Technology and Engineering, Vellore Institute of Technology, Vellore, Tamil Nadu, India.  She completed her Doctoral degree in Vellore Institute of Technology, Vellore. Her areas of interest are Machine Learning, Soft computing, Data mining, Artificial Intelligence, Predictive Analytics.
\end{IEEEbiographynophoto}
\vskip -2\baselineskip 
\begin{IEEEbiographynophoto}{Quoc-Viet Pham} [M'18] (vietpq@pusan.ac.kr) is currently working as a research professor at Korean Southeast Center for the 4th Industrial Revolution Leader Education, Pusan National University, Korea. He has been granted the Korea NRF Funding for outstanding young researchers for the term 2019–2023. He received the best PhD thesis award in Engineering from Inje University in 2017. His research interests include network optimization, edge computing, resource allocation, and wireless AI.
\end{IEEEbiographynophoto}
\vskip -2\baselineskip 
\begin{IEEEbiographynophoto}{Dinh C. Nguyen}
is currently pursuing the Ph.D. degree at Deakin University, Australia. His research interests focus on blockchain, reinforcement learning and edge computing. He received a prestigious Data61 Ph.D. scholarship, Australia.
\end{IEEEbiographynophoto}
\vskip -2\baselineskip 
\begin{IEEEbiographynophoto}{Praveen Kumar Reddy M} received Ph.D. degree from VIT, Vellore, India. He had worked as a Software Developer with IBM, Alcatel-Lucent. He is currently working as an Assistant Professor with the School of Information Technology and Engineering, VIT. His research interests include energy-aware applications for the Internet of Things (IoT) and high-performance computing.
\end{IEEEbiographynophoto}
\vskip -2\baselineskip 
\begin{IEEEbiographynophoto}{Thippa Reddy G}
received Ph.D. degree from VIT, Vellore, India. He is currently working as an Associate Professor with the School of Information Technology and Engineering, VIT. His research interests include machine learning, deep neural networks, the Internet of Things, and blockchain.
\end{IEEEbiographynophoto}
\vskip -2\baselineskip 
\begin{IEEEbiographynophoto}{Pubudu N. Pathirana}
is a full Professor and the Director of Networked Sensing and Control group at the School of Engineering, Deakin University, Geelong, Australia. He was a visiting professor at Yale University in 2009. His current research interests include Bio-Medical assistive device design, mobile/wireless networks, and rehabilitation robotics.
\end{IEEEbiographynophoto}
\vskip -2\baselineskip 
\begin{IEEEbiographynophoto}{Octavia A Dobre} 
	(M'05-SM'07-F'20) is a Professor and Research Chair at Memorial University, Canada. Her research interests include technologies for beyond 5G, as well as optical and underwater communications. She published over 300 referred papers in these areas. Dr. Dobre serves as the Editor-in-Chief (EiC) of the IEEE Open Journal of the Communications Society. She was the EiC of the IEEE Communications Letters, a senior editor and an editor with prestigious journals, as well as General Chair and Technical Co-Chair of flagship conferences in her area of expertise. Dr. Dobre is the recipient of diverse awards, such as Best Paper Awards at IEEE ICC, IEEE Globecom, and IEEE WCNC conferences. She is a Distinguished Lecturer of the IEEE Communications Society and a fellow of the Engineering Institute of Canada.
\end{IEEEbiographynophoto}
\end{document}